\documentclass[conference]{IEEEtran}
\IEEEoverridecommandlockouts

\usepackage{cite}
\usepackage{amsmath,amssymb,amsfonts}
\usepackage{algorithmic}
\usepackage{graphicx}
\usepackage{textcomp}
\usepackage{stfloats}

\usepackage{balance}
\usepackage{flushend}
\flushend
\usepackage{booktabs} 
\usepackage{diagbox}
\usepackage{hyperref}
\usepackage{xcolor}
\usepackage{tabularx}

\def\BibTeX{{\rm B\kern-.05em{\sc i\kern-.025em b}\kern-.08em
    T\kern-.1667em\lower.7ex\hbox{E}\kern-.125emX}}
\begin{document}

\title{Lungmix: A Mixup-Based Strategy for Generalization in Respiratory Sound Classification
\thanks{We would like to thank Shenzhen HealAll Technology Co., Ltd for providing medical resources for this research. Company's website: \url{https://healall.net/}}
}

\author{
\IEEEauthorblockN{Shijia Ge, Weixiang Zhang, Shuzhao Xie, Baixu Yan, Zhi Wang}
\IEEEauthorblockA{\textit{Shenzhen International Graduate School} \\
\textit{Tsinghua University}\\
Shenzhen, China \\
\{gsj23, zhang-wx22, ybx22, xsz24\}@mails.tsinghua.edu.cn, wangzhi@sz.tsinghua.edu.cn}
}

\maketitle

\begin{abstract}
Respiratory sound classification plays a pivotal role in diagnosing respiratory diseases. While deep learning models have shown success with various respiratory sound datasets, our experiments indicate that models trained on one dataset often fail to generalize effectively to others, mainly due to data collection and annotation \emph{inconsistencies}. To address this limitation, we introduce \emph{Lungmix}, a novel data augmentation technique inspired by Mixup. Lungmix generates augmented data by blending waveforms using loudness and random masks while interpolating labels based on their semantic meaning, helping the model learn more generalized representations. Comprehensive evaluations across three datasets, namely ICBHI, SPR, and HF, demonstrate that Lungmix significantly enhances model generalization to unseen data. In particular, Lungmix boosts the 4-class classification score by up to 3.55\%, achieving performance comparable to models trained directly on the target dataset.

\end{abstract}


\section{Introduction}

Lung auscultation \cite{NEJLung} is the most commonly used method in diagnosing respiratory diseases, in which the recognition of respiratory sound plays a critical role. This method is exceptionally rapid, cost-effective, hygienic, and notably safe, setting it apart from other diagnostic tools like Computed Tomography (CT) or Magnetic Resonance Imaging (MRI) scans. In general, only experienced physicians can accurately diagnose diseases based on breath sounds. This expertise requirement poses a significant challenge to the widespread availability of affordable stethoscope services, particularly in remote areas with limited medical resources. 

The invention of electronic stethoscopes enables the collection of many respiratory sound datasets \cite{ICBHI, SPRSound2022, HF} to train machine learning models, making low-cost and manual-free respiratory disease diagnosis practical. Early methods using traditional approaches \cite{lungsoundsurvey} and convolutional neural networks (CNNs) \cite{LungRN, respirenet, AdventitiousRC,cnnmoe} showed some improvement in respiratory sound classification but were outperformed by the more powerful Audio Spectrogram Transformers (AST) \cite{AST}. Many approaches \cite{patchmix, repaug, BTS, M2d} effectively harness Transformer-based models to process spectral representations of respiratory sound signals, achieving remarkable performance.

\begin{figure}
    \centering
    \includegraphics[width=0.99\linewidth]{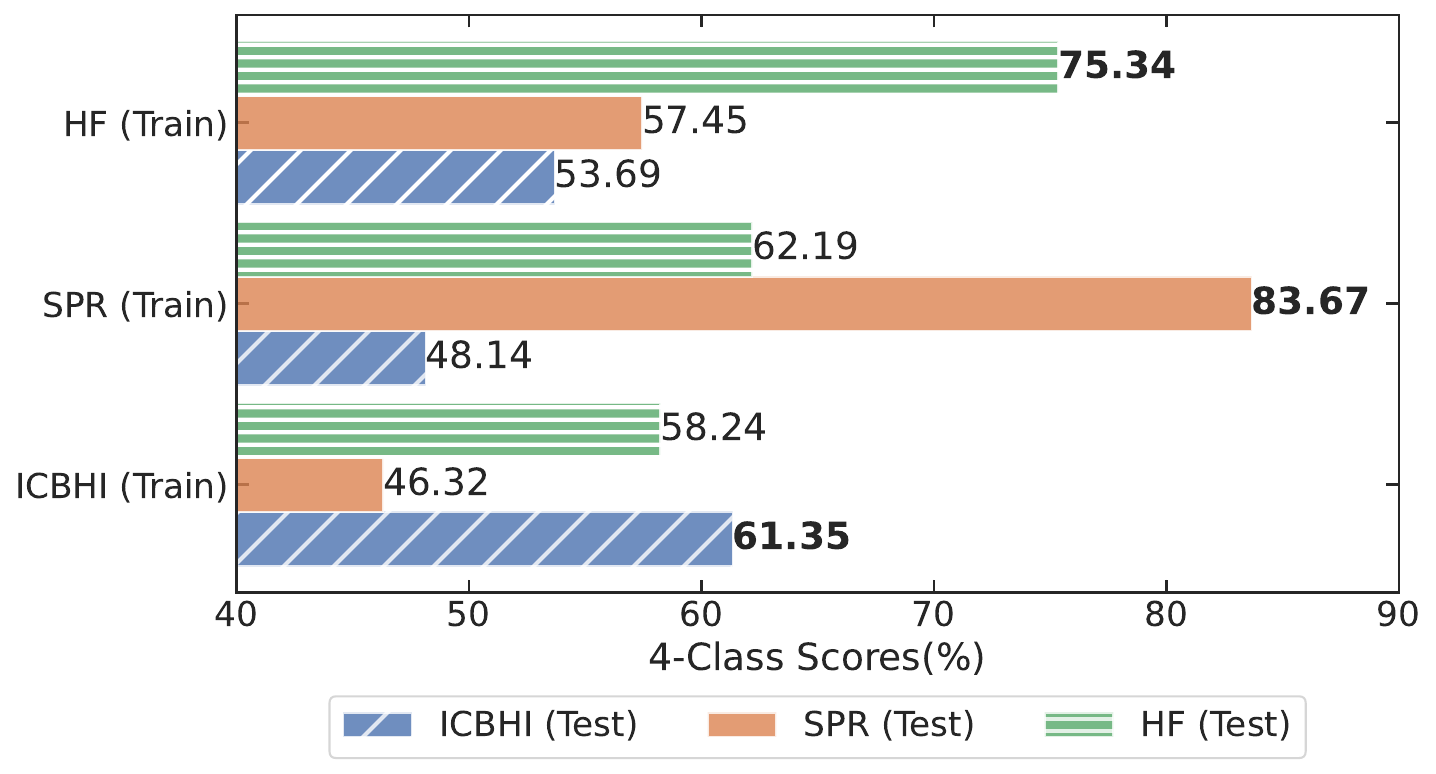}
    \caption{Performance comparison of simple fine-tuned audio spectrogram transformers. While the models demonstrated strong performance on their respective test datasets, their accuracy significantly deteriorated when evaluated on the two unseen datasets, with a performance degradation of more than 30\% in some cases. The calculation of scores and dataset descriptions are detailed in the experiment section.}
    \label{fig:moti}
    \vspace*{-16.6pt}
\end{figure}

Despite these advancements, most existing works are limited to paying little attention to enhancing the domain generalization of these models. Only a few works \cite{kim2023stethoscopeguidedsupervisedcontrastivelearning, DomainTransferAug} consider domain shift between the train and test data inside one dataset. As a result, these models often struggle to adapt to unseen domains. \autoref{fig:moti} illustrates this challenge, where models trained on three distinct datasets exhibit a considerable drop in scores when tested on unseen datasets. This poor generalization presents the primary barrier to the widespread adoption of deep learning methods for lung auscultation.

One straightforward method to enhance model generalization is to increase the amount of training data. However, collecting new respiratory sound data is costly and complex. To address this issue, we propose a novel data augmentation method based on Mixup \cite{Mixupemorigin}, a well-known method that could generate additional data and improve generalization \cite{zhang2021doesmixuphelprobustness} efficiently. However, previous applications of Mixup to respiratory sounds have always followed the assumption that the interpolation of labels obeys a linear relationship. While standard Mixup has been successfully applied to CNN-based models \cite{LungRN, respirenet, GAPAUG}, it does not work well with transformers \cite{patchmix} due to the standard \(\lambda\)-interpolation. We further investigate and verify the relationship between label interpolation and model performance.

Additionally, we identify another key challenge with existing Mixup approaches in respiratory sound classification. Abnormal respiratory sounds like \textbf{\textit{wheeze}} and \textit{\textbf{crackle}} often occur sparsely within an audio signal. Randomly mixing signals with standard Mixup can not produce samples that accurately capture the key features of these abnormal sounds, making it challenging to create training data aligned with their labels. We conduct experiments with the previous state-of-the-art method Patch-mix \cite{patchmix} to validate this hypothesis.

Building on these findings, we introduce \emph{Lungmix}, a novel Mixup-based approach tailored for respiratory sound classification. Unlike traditional methods, Lungmix integrates the semantic meaning of labels and multi-label classification into the label interpolation process. Besides, Lungmix is applied to waveforms rather than mel spectrogram. We use a mask derived from the waveform's loudness to generate more plausible data, which preserves more fine-grained information and unburdens the need for additional models \cite{kim2020puzzlemixexploitingsaliency, qin2020resizemix, AttentionMixup}. 
Consequently, Lungmix enhances the generalization of models across diverse datasets while retaining crucial audio features.

Our contributions can be summarized as follows. (1) We pinpoint the core issue of poor generalization in respiratory sound classification and demonstrate it through carefully designed experiments. (2) We introduce a data augmentation technique for respiratory sound classification. It leverages loudness masks and random masks to create new data samples and interpolates labels according to their semantic meaning. (3) We align differences between datasets, ensuring a fair and consistent evaluation of various Mixup algorithms. Our methods are tested on multiple datasets, showing strong generalization performance on unseen data. 

\section{Methods}
\label{sec:methods}
\noindent\textbf{Preliminaries}. The Mixup technique can be described as creating a convex combination of data \(x\) and their labels \(y\) with \(\lambda\), where \(\lambda \sim Beta(\alpha,\alpha)\), for \(\alpha \in (0, \inf)\):
\setlength{\arraycolsep}{2pt}
\begin{eqnarray}
\tilde{x} &=& \lambda \cdot x_i + (1 - \lambda) \cdot x_j,\\
\tilde{y} &=& \lambda \cdot y_i + (1 - \lambda) \cdot y_j.
\end{eqnarray}
\setlength{\arraycolsep}{5pt}

\subsection{Overview of Lungmix}
\begin{figure}
    \centering
    \includegraphics[width=0.99\linewidth]{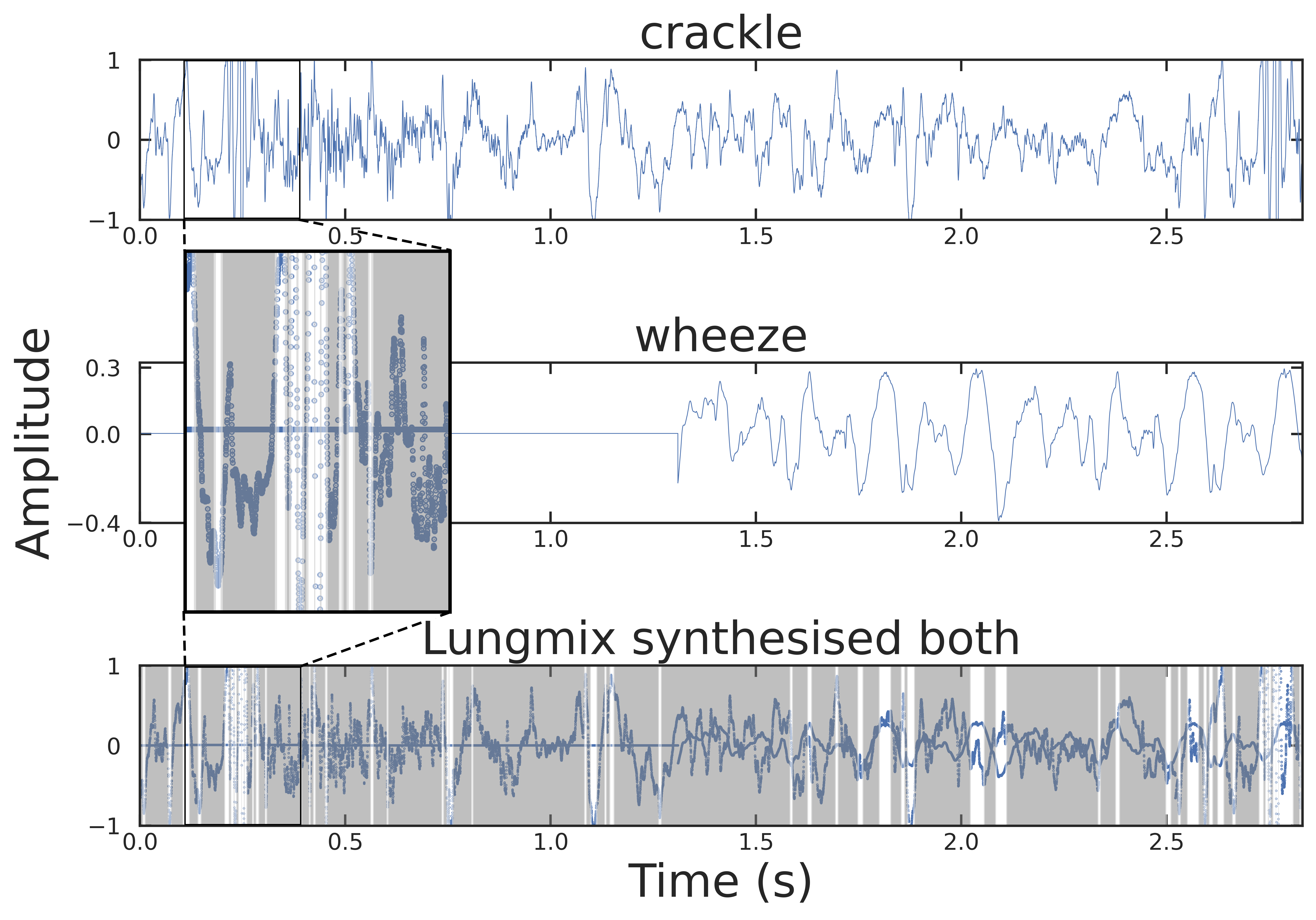}
    \caption{Visualization of Lungmix. An a \textbf{\textit{crackle}} and \textbf{\textit{wheeze}} are mixed into \textbf{\textit{both}}. The grey parts denote the random mask, and the white parts denote the loudness mask. The zoomed-in section highlights the short and discontinuous crackle sound. The part under the zoomed-in is randomly generated padding.}
    \label{fig:wav}
    \vspace*{-6pt}
\end{figure}

Our goal is to train a model on one dataset but generalize well to others. This can be formulated as a single-source domain generalization (SSDG) problem in the context of respiratory sound classification, as outlined in \cite{dgsurvey}. We can only access respiratory sound data from one source domain \( \mathcal{S} \) in this scenario. \( \mathcal{S} \) is defined by joint distribution \( P^{(S)}_{\mathcal{X}\mathcal{Y}} \) where \( \mathcal{X} \) represents the input space and \( \mathcal{Y} \) represents the target space. The objective is to train a predictive model \( g: \mathcal{X} \rightarrow \mathcal{Y} \) to minimize the empirical risk on this source domain while ensuring robust performance across unseen target domains \( T_i \). Each target domain \( T_i \) has its own joint distribution \( P^{(i)}_{\mathcal{X}\mathcal{Y}} \).

To this end, we designed the Lungmix method, which improves the generalization of deep-learning models trained for respiratory disease diagnosis through data augmentation. The Lungmix is defined as follows:
\setlength{\arraycolsep}{2pt}
\begin{eqnarray}
\tilde{x} &=& \textbf{M}_{\text{Lungmix}} \odot x_{i}+(1-\textbf{M}_{\text{Lungmix}}) \odot x_{j},\\
\tilde{y} &=& y_i \oplus y_j.
\end{eqnarray}
\setlength{\arraycolsep}{5pt}
\setlength{\arraycolsep}{5pt}
It combines two waveforms \(x_i\) and \(x_j\) using a mask  \(\textbf{M}_{\text{Lungmix}} \in \{0, \lambda, 1\}^{1 \times maxlen(x_i,x_j)}\) comes from the random mask and the loudness mask, \(\odot\) is element-wise multiplication. While the labels \(y_i\) and \(y_j\) are interpolated through bitwise OR operation \(\oplus\) to calculate the final mixed label.

\subsection{Mask Generation}

Let \(x_i\) and \(x_j\) be two audio waveforms that will be mixed. First, we apply random shifting and rolling to one of the two waveforms to generate diverse data structures. Then, we separate the meaningful parts of the waveform according to their loudness. Loudness masks \(\textbf{M}_i\) and \(\textbf{M}_j\) are created for \(x_i\) and \(x_j\) using the standard variance to separate outliers. The final loudness mask \(\textbf{M}_{\text{Loudness}}\) is combined by bitwise OR and then multiplied by \(\lambda\), where lambda belongs to Beta distribution just the way in the vanilla Mixup. Each element of the meaningful part is combined using the loudness mask drawn from, thus preserving features from both audio waveforms.
\setlength{\arraycolsep}{2pt}
\begin{eqnarray}
 \textbf{M}_i &=& |\mathbf{x_i}| > |\text{mean}(x_i) + 2 \cdot \text{std}(x_i)|,\\
\textbf{M}_{\text{Loudness}} &=& \lambda \cdot (\textbf{M}_i\oplus \textbf{M}_j).
\end{eqnarray}
\setlength{\arraycolsep}{5pt}

After that, we combine the remaining parts using a binary random mask \(\textbf{R}\) according to 
\setlength{\arraycolsep}{2pt}
\begin{equation}
\textbf{R}[i,j] = \left\{
\begin{IEEEeqnarraybox}[\relax][c]{l's}
1, & \text{for } $\mathcal{U}(0, 1) > 0.5$ \\
0, & \text{otherwise}
\end{IEEEeqnarraybox}
\right.
\label{eq:4}
\end{equation}
\setlength{\arraycolsep}{5pt}
where each bit is determined by comparing a uniform random variable belonging to uniform distribution \(\mathcal{U}(0, 1)\).
In the end, we calculate the final mask, where \(\oplus\) is element-wise OR operation to combine two masks:
\setlength{\arraycolsep}{2pt}
\begin{equation}
\begin{aligned}
\textbf{M}_{Lungmix}&=\textbf{M}_{Loudness} \oplus\textbf{R}.
\end{aligned}
\label{eq:5}
\end{equation}
\setlength{\arraycolsep}{5pt}


\subsection{Label Interpolation}

\begin{figure}
    \centering
    \includegraphics[width=\linewidth]{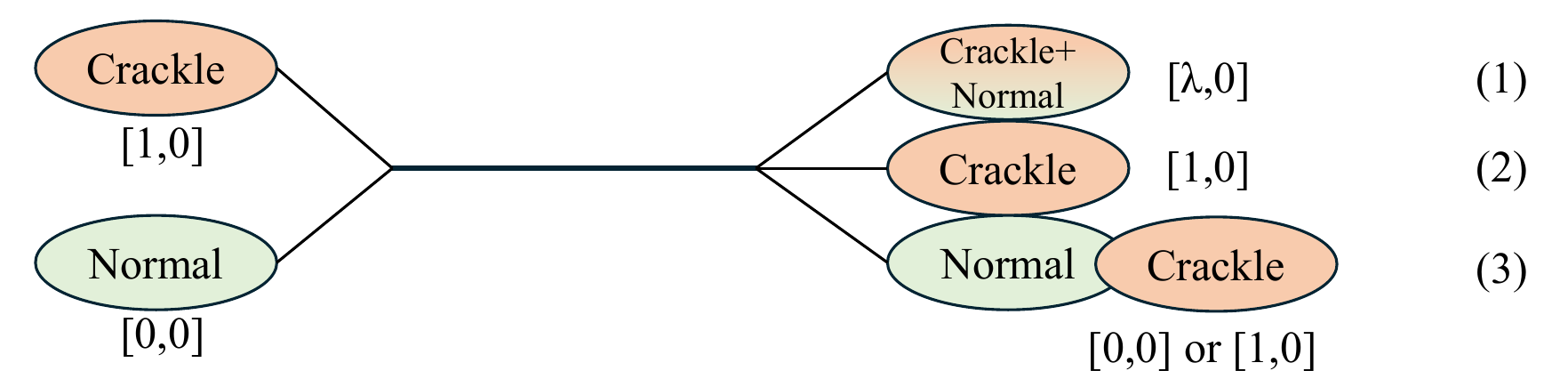}
    \caption{Visualization of label interpolation. (1) is linear interpolation, (2) is non-linear interpolation, (3) is label preservation.}
    \label{fig:lin}
    \vspace*{-10pt}
\end{figure}
In the context of respiratory sound classification, assume we have \( n \) distinct types of respiratory sounds, where we designate \( \text{class}_0 \) as normal respiratory sounds and the remaining \( n-1 \) classes as abnormal respiratory sounds. In practical scenarios, it is possible to have combinations of \( 2 \) to \( n-1 \) abnormal respiratory sounds, resulting in \( C_{n-1}^2 + C_{n-1}^3 + \cdots + C_{n-1}^{n-1} = 2^{n-1} - n + 1 \) different categories of mixed abnormal respiratory sounds. Our approach, in essence, handles this problem as a multi-label classification task, where each distinct combination of abnormal sounds can be viewed as a separate label. By adopting a Label Powerset (LP) approach \cite{Tsoumakas2010MiningMD}, we transform these multi-label combinations into distinct categories, treating each possible combination of abnormal sounds as a unique class. 

In our study, we assume that when any normal respiratory sound (\( \text{class}_0 \)) is mixed with abnormal respiratory sounds, the resulting category retains the original abnormal respiratory sound category. To systematically handle the classification of mixed respiratory sounds, for combination \( n \) respiratory sound signals \( y_1, y_2, \dots, y_n \), we first transform the class labels into one-hot encoded vectors of length \(n-1\), where each bit represents an abnormal sound class. Once these one-hot vectors are created, we apply a bitwise OR operation (\(\oplus\)) to combine multiple respiratory sounds. By doing so, we can capture combinations of abnormal sounds and then map them to one of \(2^{n-1} + 1\) distinct categories (including the normal class). Thus, turn the multi-label task into a multi-class task.

We also consider another possible scenario where Mixup does not alter the labels of the mixed inputs, which we refer to as \textbf{\textit{label preservation}}. In this case, the label is determined by the random selection of the data being mixed. In our implementation, we select the label of the first input. We explore this method because, even if the label is slightly incorrect, it may still help the model learn effectively \cite{Wei2020TheoreticalAO}.

In the context of a four-class respiratory sound classification problem, where the abnormal categories are limited to \textbf{\textit{crackle}} and \textbf{\textit{wheeze}}, the mixed category is defined as \textbf{\textit{both}}, indicating the presence of a mixed abnormal respiratory sound that includes both \textbf{\textit{crackle}} and \textbf{\textit{wheeze}}. \(\mathcal{L}_{CE}\) denotes the classic cross-entropy loss. The typical Mixup cross-entropy loss \(\mathcal{L}_{\text{Mixup}}\) is as follows: 
\setlength{\arraycolsep}{2pt}
\begin{equation}
\begin{aligned}
\mathcal{L}_{\text{Mixup}} &= \lambda \cdot \mathcal{L}_{CE}(g(\tilde{x}), y_1)
\\&\quad+ (1 - \lambda) \cdot \mathcal{L}_{CE}(g(\tilde{x}), y_2).
\end{aligned}
\label{eq:Mixuploss}
\end{equation}

However, this loss format is not suitable for respiratory sounds. The reason is that once an abnormal audio sample is mixed with a normal one, it is still considered entirely abnormal. For example, a mix of \textbf{\textit{normal}} and \textbf{\textit{crackle}} sounds would still be labeled as \textbf{\textit{crackle}} rather than a mixture (e.g., 0.4 * \textbf{\textit{normal}} + 0.6 * \textbf{\textit{crackle}}). Furthermore, we adjust the degree of non-linearity in our label interpolation by combining traditional linear loss to provide regularization to models. Our loss function can be written as in \eqref{eq:2}. During the experiment, we set \(\lambda_{1}\) to 1, and \(\lambda_{2}\) is rescaled \(\mathcal{L}_{\text{Mixup}}\) according to the first loss:
\setlength{\arraycolsep}{2pt}
\begin{equation}
\begin{aligned}
\mathcal{L} &= \lambda_{1} \cdot \mathcal{L}_{\text{CE}}(g(\tilde{x}), \tilde{y})+ \lambda_{2} \cdot \mathcal{L}_{\text{Mixup}}.
\end{aligned}
\label{eq:2}
\end{equation}
\setlength{\arraycolsep}{5pt}

\section{Experiment and Results }

\subsection{Dataset Description}
We evaluated our methods on three different respiratory sound datasets: \textbf{(1) ICBHI} \cite{ICBHI}: It is segmented based on respiratory cycles and annotated for the presence of crackles and wheezes, accompanied by detailed metadata. \textbf{(2) SPRSound} \cite{SPRSound2022}: It annotates at record and event levels with seven class labels. We use event level in our experiment. \textbf{(3) HF} \cite{HF}: It shares similarities with SPRSound but introduces additional labels for inhale and exhale phases alongside four abnormal respiratory sound classes. However, it lacks annotations for segments containing both crackles and wheezes.

We used the official train and test splits for a fair comparison across these datasets. During preprocessing, the labels were unified into a standard four-class system based on ICBHI: \textbf{\textit{normal}}, \textbf{\textit{crackle}}, \textbf{\textit{wheeze}}, and \textbf{\textit{both}}. Discontinuous, non-musical sound classes such as\textbf{\textit{ coarse crackle}} and \textbf{\textit{fine crackle}} were categorized as \textbf{\textit{crackle}}, while continuous, musical sound classes like \textbf{\textit{stridor}} and \textbf{\textit{rhonchus}} were reclassified as \textbf{\textit{wheeze}}.

\subsection{Experiments Setup}
\textbf{Preprocessing Details}. We resample all the audios to 16 kHz and apply a bandpass filter to retain frequencies between 50 Hz and 1500 Hz. Also, each waveform segment is padded or cut to 9 seconds and corresponds to a 128-dimension spectrogram with 1024 frames. We also apply normalization to spectrograms with the mean and standard variation of AudioSet\cite{audioset}. \textbf{Traing Settings}. AST \cite{AST} is used as the experiment's classifier to compare with state-of-the-art methods. The model is pre-trained on ImageNet \cite{russakovsky2015imagenet} and AudioSet \cite{audioset}, providing a strong ability for transfer learning. Also, we use the mean of all the hidden states for classification instead of the first two tokens in the origin implementation. And we add a batch normalization \cite{batchnorm} layer before the classifier. The learning rate was set to \(1e^{-5}\), and AdamW \cite{adamw} was used as an optimizer to fine-tune the model. Also, cosine annealing  \cite{cosine} was used as a learning rate scheduler. Besides, we use a batch size of 4 with a gradient accumulation of 8 to achieve the equivalence of a batch size of 32. 
\textbf{Evaluation Metric}. We adopted the evaluation metric from the ICBHI 2017 challenge to enable a fair comparison. The challenge defined the Average Score (Sc) as the average of Sensitivity (Se) and Specificity (Sp). Each term is defined as follows: \( C_N \), \( C_C \), \( C_W \), and \( C_B \) represent the number of samples correctly classified as \textbf{\textit{normal}}, \textbf{\textit{crackle}}, \textbf{\textit{wheeze}}, and \textbf{\textit{both}}, respectively. In contrast, \( N_N \), \( N_C \), \( N_W \), and \( N_B \) denote the total number of samples for the corresponding class. The metrics are defined as follows: 

\begin{equation}
\begin{aligned}
Se=\frac{C_C+C_W+C_B}{N_C+N_W+N_B},Sp=\frac{C_N}{N_N},Sc=\frac{Se+Sp}{2}.
\end{aligned}
\label{eq: metric}
\end{equation}

\subsection{Comprehensive Result}
\begin{table*}
\centering
\caption{The score for the AST model trained using various Mixup methods and source domains. Higher is Better.}
\label{tab: 0}
\begin{tabularx}{\textwidth}{@{}l|XXXX|XXXX|XXXX@{}}
\hline
Source Domain & \multicolumn{4}{c|}{ICBHI} & \multicolumn{4}{c|}{SPR} & \multicolumn{4}{c}{HF} \\ 
\hline
Test Domain&ICBHI&SPR&HF&COMB&ICBHI&SPR&HF& 
COMB&ICBHI&SPR&HF&COMB\\
\hline\hline
w/o Mixup&52.88&65.48&61.71&60.79&\underline{48.9}&80.44&59.29&62.52&\underline{58.24}&	62.19&75.34&68.56
\\
vanilla Mixup\cite{Mixupemorigin}&
\textbf{58.53}&61.27&60.62&60.53&
47.16&53.85&53.85&57.86&
\textbf{58.92}&62.67&75.56&68.74
\\
Cutmix\cite{cutmix}&55.53&	64.72&	62.98&	61.82&45.19&77.65&54.37	&58.76&58.15&63.41&73.89&68.66
\\
Patch-mix\cite{patchmix}&
50.32&61.76&59.48&58.26&
46.48&\underline{80.76}&57.43&60.91&
56.08&56.89&72.03&65.57
\\
\hline\hline
Patch-mix (label preservation)&
\underline{57.41}&61.83&\underline{64.86}&62.08&
46.84&\textbf{81.17}&59.59&62.40&
55.50&67.72&75.96&70.32
\\
Lungmix (w/o loudness)&
\textbf{58.53}&61.26&60.32&59.46&
42.35&76.27&55.67&58.80&
56.20&67.72&74.05&70.90
\\
Lungmix (linear)&
52.13&\underline{67.04}	&60.69&60.94&
48.46&77.58&\underline{61.49}&\underline{62.70}&
52.54&63.58&77.53&69.12
\\
Lungmix (combined)&
53.78&\textbf{68.59}&60.52&61.71&
50.13&80.21&56.30&61.03&
56.46&62.61&74.95&68.86
\\
\textbf{Lungmix (non-linear)}& 
53.35&66.63&\textbf{65.00}&\textbf{63.30}&
\underline{50.56}&80.61&53.93&	59.93
&56.37&\underline{67.81}&\underline{77.35}&\textbf{72.11}
\\
\textbf{Lungmix + Patch-mix}& 
52.82&66.65&63.52&\underline{63.06} &
\textbf{51.40}&79.06&\textbf{62.12}&\textbf{64.07}&
55.75&\textbf{69.37}&\textbf{77.76}&\underline{72.08}
\\
\hline
\multicolumn{13}{@{}X}{
  \begin{minipage}{16cm} The \textbf{bolded results} represent the best performance, and the \underline{underlined results} represent the second-best performance in the target domains.
  \end{minipage}
}\\
\end{tabularx}
\vspace*{-4pt}
\end{table*}
\autoref{tab: 0} summarizes the performance of the AST models trained using various Mixup methods across three source domains: ICBHI, SPR, and HF. The performance is also evaluated on the combined test data, denoted as COMB, to simulate real-world distribution. The top half of the table presents previous methods, while the bottom half details our proposed methods. The AST model trained without Mixup augmentation is the baseline, achieving moderate performance across all domains. Other traditional Mixup techniques fall short of addressing the issue of domain generalization, with almost unimproved scores or sometimes drops.

The results demonstrate that Lungmix is the most effective method, significantly enhancing performance across the different source and test domains. Even though results are slightly inverted when using SPRSound as the source domain, Lungmix still achieves the best overall performance. When Lungmix is applied without the loudness mask (using only a random mask), no improvement is observed, highlighting the importance of the loudness-based mask in the process.

We also identify why Patch-mix initially underperforms. In the row labeled Patch-mix and ours modified Patch-mix (label preserved), we do not apply mixup loss to the Patch-mix inputs but instead use the original labels. This approach yields much better results, supporting the hypothesis that features in the spectrogram are sparse and Patch-mix does not significantly alter input labels. The result also confirms that Patch-mix is an effective regularization tool. We combine Lungmix and Patch-mix, which results in the best performance on the SPR and HF datasets.

Notice that the result for training and testing on ICBHI is not as good as in many previous methods, mainly because we are not using the best result on ICBHI. Instead, we chose the best result on the COMB dataset. This kind of trade-off shows the sacrifice required to increase generalization capability. 

What is more, a key observation is that, despite the HF dataset lacking a \textbf{\textit{both}} class, models trained on HF achieve the best overall score, reaching 72.11\% and outperforming the baseline by 3.55\%. Moreover, the performance on ICBHI, when trained on HF, is close to that of models trained directly on ICBHI. Despite the HF test dataset being larger than those of ICBHI and SPR, these results still underscore the superior quality of the HF data compared to the other datasets.


\subsection{Ablation Study for Non-linear Interpolation for Mixup}

\begin{table}
\centering
\caption{Sp and Se on COMB dataset. Higher is Better}
\label{tab:augmentation}
\begin{tabular}{lllllll}
\hline
Source Domain & \multicolumn{2}{c}{ICBHI} & \multicolumn{2}{c}{SPR} & \multicolumn{2}{c}{HF} \\
\hline
Method & Sp(\%) & Se(\%) & Sp(\%) & Se(\%) & Sp(\%) & Se(\%) \\
\hline\hline
linear & 60.21&	\textbf{61.67}& \textbf{75.70} & 49.70 & 76.12 & 62.12 \\
combined & 70.96&52.47& 65.32 & \textbf{56.74} & \textbf{81.41} & 56.30 \\
non-linear & \textbf{79.08} & 47.51 & 63.64 & 56.22 & 77.17 & \textbf{67.04} \\
\hline
\end{tabular}
\vspace*{-4pt}
\end{table}


We conducted an ablation study to investigate whether the interpolation should be non-linear and based on the semantic meaning of the data. As shown in \autoref{tab:augmentation}, during the data augmentation, Mixup synthesizes more difficult-to-learn like \textbf{\textit{both}}, which should have enabled effective learning of representations for different classes. However, non-linear interpolation does not significantly enhance specificity or sensitivity, but it increases the score when applied to ICBHI and HF.







\section{Conclusion}

\label{sec:conclusion}

In this study, we proposed a new Mixup-based method called Lungmix. This method improves the model's generalization capabilities through data augmentation techniques, allowing it to perform better on different respiratory sound datasets. Specifically, Lungmix employs a loudness-based mask mechanism for mixing operations, outperforming traditional Mixup methods across multiple datasets. Experimental results demonstrate that Lungmix effectively enhances the model's performance on unseen datasets, thereby validating its potential in addressing domain generalization issues. Future work will explore additional data augmentation strategies and technologies to enhance further the accuracy and practicality of respiratory sound classification models.

\bibliographystyle{IEEEtran}
\bibliography{ref/reference}

\end{document}